\documentclass[,11pt,letterpaper,twocolumn]{article}

\usepackage{iftex}
\ifPDFTeX
  \usepackage[T1]{fontenc}
  \usepackage[utf8]{inputenc}
  \usepackage{textcomp}
  \usepackage{lmodern}
  \DeclareUnicodeCharacter{2212}{-}
  \DeclareUnicodeCharacter{2013}{--}
  \DeclareUnicodeCharacter{2014}{---}
  \DeclareUnicodeCharacter{2011}{-}
  \DeclareUnicodeCharacter{2019}{'}
  \DeclareUnicodeCharacter{201C}{``}
  \DeclareUnicodeCharacter{201D}{''}
  \DeclareUnicodeCharacter{2192}{\ensuremath{\to}}
  \DeclareUnicodeCharacter{2264}{\ensuremath{\le}}
  \DeclareUnicodeCharacter{2265}{\ensuremath{\ge}}
  \DeclareUnicodeCharacter{0394}{\ensuremath{\Delta}}
  \DeclareUnicodeCharacter{2208}{\ensuremath{\in}}
\else
  \usepackage{unicode-math} 
  \defaultfontfeatures{Scale=MatchLowercase}
  \defaultfontfeatures[\rmfamily]{Ligatures=TeX,Scale=1}
\fi
\IfFileExists{microtype.sty}{\usepackage[]{microtype}\UseMicrotypeSet[protrusion]{basicmath}}{}

\usepackage[margin=0.7in]{geometry}
\usepackage{graphicx}
\graphicspath{{reports/}{reports/figures/}{figures/}{./}}
\usepackage{longtable,booktabs,array}
\usepackage{float}
\usepackage{tabularx}
\usepackage{etoolbox}

\usepackage{amsmath,amssymb}
\usepackage{enumitem}
\usepackage{fvextra}
\fvset{
  fontsize=\footnotesize,
  breaklines=true,
  breakanywhere=true,
  xleftmargin=0pt,
  numbers=none,
  breaksymbolleft={}
}
\usepackage{csquotes}
\usepackage{calc}
\usepackage{placeins}
\usepackage{caption}

\usepackage{xcolor}
\definecolor{ArXivLink}{HTML}{0B3D91}
\usepackage{hyperref}
\hypersetup{
  colorlinks=true,
  linkcolor=black,
  citecolor=ArXivLink,
  urlcolor=ArXivLink,
  pdfauthor={Travon Lucius, Contributor Name, Contributor
Name, Contributor Name, Contributor Name},
  pdftitle={Deep Hedging with Reinforcement Learning: A Practical
Framework for Option Risk Management}
}

\makeatletter
\newsavebox\pandoc@box
\newcommand*\pandocbounded[1]{
  \sbox\pandoc@box{#1}%
  \Gscale@div\@tempa{\textheight}{\dimexpr\ht\pandoc@box+\dp\pandoc@box\relax}%
  \Gscale@div\@tempb{\linewidth}{\wd\pandoc@box}%
  \ifdim\@tempb\p@<\@tempa\p@\let\@tempa\@tempb\fi
  \ifdim\@tempa\p@<\p@\scalebox{\@tempa}{\usebox\pandoc@box}%
  \else\usebox{\pandoc@box}%
  \fi%
}
\def\fps@figure{htbp}
\makeatother

\makeatletter
\renewcommand{\maketitle}{
  \begin{center}
    {\LARGE\bfseries Deep Hedging with Reinforcement Learning: A
Practical Framework for Option Risk Management\par}\vspace{0.5em}
    {\small
      \begin{tabular}{@{}l@{}}
      Travon Lucius $\mid$ \texttt{travon.lucius@blackrock.com}\\
      Christian (Chip) Koch Jr $\mid$ \texttt{chipgad@gmail.com}\\
      Jacob Starling $\mid$ \texttt{jacobstarling4313@gmail.com}\\
      Julia Zhu $\mid$ \texttt{julia20001024@gmail.com}\\
      Miguel Urena $\mid$ \texttt{miguel.urena@emory.edu}\\
      Carrie Hu $\mid$ \texttt{carrie.hu@emory.edu}\\
      \end{tabular}\par}
    \vspace{0.25em}{\normalsize November 2025\par}
  \end{center}
}
\makeatother

\newenvironment{arxivabstract}{
  \small
  \begin{center}\bfseries Abstract\end{center}
}{\par\vspace{0.5em}}

\usepackage{titlesec}
\titleformat{\section}{\large\bfseries}{\thesection.}{0.5em}{}
\titleformat{\subsection}{\normalsize\bfseries}{\thesubsection}{0.5em}{}

\NewDocumentCommand\citeproctext{}{}
\NewDocumentCommand\citeproc{mm}{%
  \begingroup\def\citeproctext{#2}\cite{#1}\endgroup}
\makeatletter
 \let\@cite@ofmt\@firstofone
 \def\@biblabel#1{}
 \def\@cite#1#2{{#1\if@tempswa , #2\fi}}
\makeatother
\newlength{\cslhangindent}
\newlength{\csllabelwidth}
\newenvironment{CSLReferences}[2] 
 {\begin{list}{}{%
  \setlength{\itemindent}{0pt}
  \setlength{\leftmargin}{0pt}
  \setlength{\parsep}{0pt}
  \ifodd #1
   \setlength{\leftmargin}{\cslhangindent}
   \setlength{\itemindent}{-1\cslhangindent}
  \fi
  \setlength{\itemsep}{#2\baselineskip}}}
 {\end{list}}

\providecommand{\tightlist}{\setlength{\itemsep}{0pt}\setlength{\parskip}{0pt}}

\begin{document}

\twocolumn[
\maketitle
\begin{arxivabstract}
We present a reinforcement-learning (RL) framework for
dynamic hedging of equity index option exposures under realistic
transaction costs and position limits. We hedge a normalized
option-implied equity exposure (one unit of underlying delta, offset via
SPY) by trading the underlying index ETF, using the option surface and
macro variables only as state information and
not as a direct pricing engine. Building on the ``deep hedging''
paradigm of Buehler et al.
(\citeproc{ref-buehler2019deephedging}{Buehler et al. 2019}), we design
a leak-free environment, a cost-aware reward function, and a lightweight
stochastic actor--critic agent trained on daily end-of-day data panel constructed from
SPX/SPY implied volatility term structure, skew, realized volatility,
and macro rate context. On a fixed train/validation/test split, the
learned policy improves risk-adjusted performance versus no-hedge,
momentum, and volatility-targeting baselines (higher point-estimate
Sharpe); only the GAE policy's test-sample Sharpe is statistically
distinguishable from zero, although confidence intervals overlap with a
long-SPY benchmark so we stop short of claiming formal dominance.
Turnover remains controlled and the policy is
robust to doubled transaction costs. The
modular codebase---comprising a data pipeline, simulator, and training
scripts---is engineered for extensibility to multi-asset overlays,
alternative objectives (e.g., drawdown or CVaR), and intraday data. From a portfolio management perspective, 
the learned overlay is designed to sit on top of an existing SPX or
SPY allocation, improving the portfolio's mean–variance trade-off with controlled 
turnover and drawdowns. We discuss practical implications for portfolio overlays and outline
avenues for future work. All code, configuration files, and experiment
scripts are available at
\url{https://github.com/tlucius16/deep-hedging-rl}.

\end{arxivabstract}

\vspace{1em}
]

\section{Executive Summary}\label{executive-summary}

\noindent We built a ``deep hedging'' stack for SPX/SPY exposures
where an RL agent learns to adjust overlay positions after seeing a
rolling window of option surface and macro features. Instead of relying
on handcrafted delta heuristics, the agent optimizes a cost-aware
reward inside a simulator and is evaluated on a fixed train/validation/test split that spans every major regime since 2005.

\medskip\noindent\textit{Key ingredients:}

\begin{description}[leftmargin=0pt,labelsep=0.65em,itemsep=0.25em,parsep=0pt]
\item[\textbf{Deterministic pipeline.}]
Raw market/volatility data flows through scripted cleaning jobs,
feature engineering, and HedgingEnv construction so experiments can be
replayed end-to-end.
\item[\textbf{Explainable agent.}]
The actor--critic policy consumes interpretable features (ATM IV, skew,
realized vol, rates) and uses a compact network (two-layer 256-unit MLP) with
squashed-Gaussian actions, making it simple to audit behavior.
\item[\textbf{Realistic execution.}]
Transaction costs, position limits, rebalance cadence, and optional
slippage functions are embedded directly in the environment to mirror
production desks.
\item[\textbf{Overlay diagnostics.}]
Beyond standalone policy metrics we measure how the RL overlay
interacts with a long-SPY sleeve, including blend-efficient frontiers,
rolling risk differentials, and year-by-year PnL attribution.
\end{description}

\medskip
Across the 2005--2023 walk-forward split the trained policy keeps
post-cost Sharpe positive, keeps the standalone overlay-equity test
drawdown within roughly −3\%, and achieves the best trade-off when 50\%
of capital is allocated to the agent and 50\% to long SPY. All artefacts such as figures, tables, and CSV
metrics, are in this repository so another team can validate or extend
the results without reverse-engineering ad-hoc notebooks. Point
estimates beat simple baselines, and only the GAE overlay's test Sharpe
is statistically distinguishable from zero, but confidence bands versus
the long-SPY benchmark remain wide enough that the edge is not
statistically bulletproof.

\section{Introduction}\label{introduction}

The traditional textbook approach to hedging options linearizes PnL via
Greeks and offsets exposures with static trades in the underlying. That
blueprint assumes frictionless execution, Gaussian shocks, and nearly
continuous rebalancing (\citeproc{ref-blackScholes1973}{Black and
Scholes 1973}; \citeproc{ref-merton1973}{Merton 1973}). Live desks
operate under the opposite conditions: spreads gap out during stress,
volatility regimes persist, and every trade incurs slippage.
Proportional costs alone can undermine delta replication
(\citeproc{ref-leland1985}{Leland 1985}), yet production overlays still
revolve around a few manually tuned rules.

Our review of SPX/SPY history highlights how fragile such rules can be.
Returns are heavy-tailed, realized volatility decays slowly, term
structure and skew swing within weeks, and macro rate shocks ripple
through option surfaces with a lag. Deep out-of-the-money (OTM) strikes
frequently drop observations, forcing strict quality filters and guarded
forward-fills before features become usable. These patterns argue for
policies that can reason over multiple steps, incorporate broader
context, and explicitly trade off hedge intensity against transaction
costs.

Reinforcement learning offers that flexibility. Rather than chase local
deltas, an agent can operate inside a simulator that embeds trading
frictions, observe a window of engineered features, and learn how
aggressively to hedge based on state. Our implementation borrows from
the ``deep hedging'' literature
(\citeproc{ref-buehler2019deephedging}{Buehler et al. 2019}) but
emphasizes reproducibility (scripted pipelines, deterministic splits)
and auditability (transparent features, compact networks).
Sections\textasciitilde{}\ref{data-and-feature-engineering}--\ref{hedging-environment}
describe the data set and environment,
Section\textasciitilde{}\ref{results} presents deterministic evaluations
including realistic execution constraints and blended overlays, and
Section\textasciitilde{}\ref{discussion-and-implications} covers
operational implications.

The figure below illustrates the macro/volatility regimes that motivate
this work, highlighting how VIX, realized volatility, and rates evolve
across the 2005--2023 window.

\noindent\textbf{Listing 1:} EDA excerpt used to summarize the macro/volatility panel.
\begin{Verbatim}
# EDA excerpt (Notebook 01)
summary = (panel[["close_spy", "vix", "rv_21d", "hvol_30d", "rate_10y"]]
              .pct_change()
              .describe(percentiles=[0.01, 0.05, 0.5, 0.95, 0.99])
              .T[["mean", "std", "min", "1%", "5%", "50%", "95%", "99%", "max"]])
\end{Verbatim}

\begin{figure}[!ht]
\centering
\includegraphics[width=\linewidth,alt={Long-horizon view of VIX, 10Y rates, realized 21d volatility, and 30d historical volatility.}]{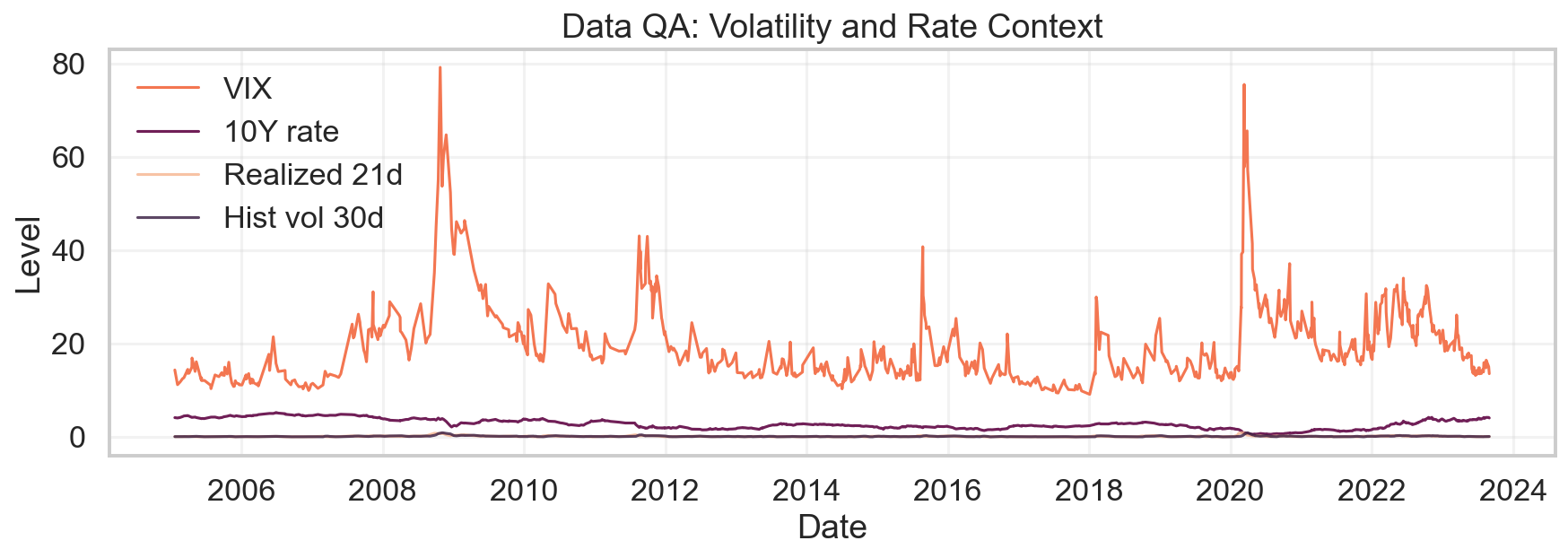}
\caption{Long-horizon view of VIX, 10Y rates, realized 21d volatility,
and 30d historical volatility.}\label{fig:exploratory}
\end{figure}

The ``deep hedging'' literature
(\citeproc{ref-buehler2019deephedging}{Buehler et al. 2019}) provides a
conceptual blueprint for that path. In this implementation we lean on
that insight but insist on practical guardrails. The simulator
eliminates look-ahead bias by aligning returns with the action timestamp,
rewards are expressed in basis points so optimization remains
numerically stable, and the policy network is intentionally compact so
behavior can be explained. The goal is not to produce an opaque '' black
box '' but to craft a framework that can be audited,
extended, and ultimately slotted into production without hidden
dependencies.

\section{Related Work}\label{related-work}

\textbf{Hedging Under Frictions.} Leland's seminal work
(\citeproc{ref-leland1985}{Leland 1985}) incorporated proportional costs
into discrete hedging and highlighted the trade-off between tracking
error and transaction cost. Later treatments extended that idea using
continuous-time control and Hamilton--Jacobi--Bellman equations
(\citeproc{ref-glasserman2004}{Glasserman 2004};
\citeproc{ref-oksendal2003}{Øksendal 2003}), but the curse of
dimensionality quickly limits those approaches once path dependence or
multiple risk factors enter the picture. Practical desks therefore fall
back on heuristics: rebalance on VIX spikes, lean into skew moves, or
fade realized/IV spreads. These rules work until they don't.

\textbf{Reinforcement Learning for Trading.} Policy-gradient and
actor--critic methods (\citeproc{ref-sutton_barto_2018}{Sutton and Barto
2018}) have been applied to order execution, market making, and
portfolio allocation (\citeproc{ref-moody1998}{Moody and Saffell 1998};
\citeproc{ref-kolm2021}{Kolm and Ritter 2021};
\citeproc{ref-zhang2020}{Zhang et al. 2020}). The common theme is to
learn a mapping from state features into actions that maximizes a
risk-adjusted reward subject to trading frictions. However, many
published examples rely on synthetic environments or omit the
engineering steps (data cleaning, leakage prevention) that make such
systems reproducible in practice.

\textbf{Deep Hedging.} Buehler et al.
(\citeproc{ref-buehler2019deephedging}{Buehler et al. 2019}) formalized
the notion of using neural policies to replicate options under cost and
risk constraints. Their experiments demonstrated that a learned policy
could outperform deltas when realistic spreads are included. The present
project adapts that blueprint but emphasizes transparency: the
environment is leak-free, the feature set is engineered rather than
latent, and every input/output artefact is versioned so the exercise can
be audited and extended.

\section{Data and Feature
Engineering}\label{data-and-feature-engineering}

We construct a daily panel for SPX/SPY from 2005 onward with the
following state features: at--the--money (ATM) implied volatility at 30d and 91d
horizons; term‑structure slope (iv\_91d − iv\_30d); 25‑delta put/call
and skew; VIX and the 10‑year Treasury yield; and realized/historical
volatilities (rv\_21d, hvol\_30d, hvol\_91d). Option features are
selected by tenor and delta tolerance with spread‑aware tie‑breaking
(src/simulator/features.py). IV series are forward‑filled only within a
small staleness window measured in calendar days
(src/simulator/build\_panel.py).
The cleaned panel spans 18{,}018 trading days; IV fields are populated
on ~96\% of rows (the remainder are dropped by spread/delta filters and
treated as missing). The 25‑delta inputs are the mid-quote IVs for the
25‑delta put and call at a 30d tenor, and skew is defined as
IV\_{25d,put} minus IV\_{25d,call}.

Target construction. Forward return $R_t$ is aligned to the action
timestamp (no look‑ahead):
\[
  R_t = \frac{P_{t+1} - P_t}{P_t}, \qquad \texttt{ret\_fwd}.
\]
The final panel combines features and ret\_fwd and is split by date
(e.g., train $\leq$ 2017‑12‑31; validation 2018--2019; test $\geq$
2020‑01‑01). Steps are scripted in data\_pipeline/ (load → standardize → build panels), and the exact feature set is recorded in
models/config.json.

We construct the daily EOD hedging panel by chaining dedicated build steps.
Because most financial time-series data are relatively unstructured, we focus 
on a compact set of economically motivated features derived from the option 
surface and macro variables. The feature set intentionally captures both cross-sectional surface information (ATM
levels, slope, skew) and macro context (VIX, rates, realized/historical
vol). Spread-aware tie-breaking favors liquid quotes and 
guarded forward-fills ensure volatility features are only propagated a
couple of calendar days before being treated as missing.

\noindent\textbf{Listing 2:} Helper for building the simulation panel and extracting state columns.

\begin{Verbatim}
from simulator.build_panel import build_sim_panel

# Compose the simulation panel with guarded forward-fills and explicit ret_fwd
panel, state_cols = build_sim_panel(
    market_df=market,
    spx_clean_dir=SPX_CLEAN,
    include_spy=False,
    ffill_limit=2, # IV features forward-fill at most 2 calendar days
)
\end{Verbatim}

Forward returns are computed as close-to-close percentage changes
shifted forward by one step, preventing the agent from seeing realized
PnL before choosing an action. Date splits (train ≤ 2017-12-31,
validation 2018--2019, test ≥ 2020-01-01) are stored in each run's
configuration, and all intermediate artefacts (cleaned snapshots, parquet
panels, scaler parameters, deterministic evaluation metrics) are
committed alongside the code.

\section{Hedging Environment}\label{hedging-environment}

We implement a deterministic environment HedgingEnv
(\texttt{src/simulator/env.py}). Each observation is a matrix of shape
$W \times F$ (window × features) normalized with statistics derived
solely from the training split. Actions are continuous hedge levels
$a_t \in [-a_{\max}, +a_{\max}]$, interpreted as the number of
underlying units hedged per unit option exposure. Transaction costs are
proportional to the absolute change in position,
$c_t = \kappa |a_t - a_{t-1}|$, where $\kappa$ is specified in basis
points. Per-step PnL is $p_t = a_t R_{t+1} - c_t$ and we scale the
reward as $r_t = 10^4 p_t$ (basis points) to keep gradients well
behaved. Episodes walk deterministically through
the panel; forward returns are aligned to the action time so there is no
look-ahead; NaNs or infinities are replaced with zeros to prevent
accidental leakage. Convenience policies in
\texttt{src/simulator/baselines.py} (no\_hedge, momentum,
volatility\_targeting, delta wrappers) are used both for sanity checks
during development and for the baseline results reported later.

To capture richer microstructure effects, we expose a
\texttt{slippage\_fn} hook that embeds the linear-quadratic price impact
models described in the \emph{Handbook of Price Impact Modeling}
(\citeproc{ref-webster2023impact}{Webster 2023}). If $q_t = a_t -
a_{t-1}$ denotes the trade size, temporary execution costs follow
$c_{\text{tmp}}(q_t) = \phi |q_t| + \tfrac{1}{2}\psi q_t^2$, while the
mid-price evolves as $\Delta S_{t+1} = \sigma \varepsilon_{t+1} +
\lambda q_t$, where $\phi$ captures spread/fees, $\psi$ the nonlinear
depth term, and $\lambda$ the permanent impact coefficient. Plugging
these expressions into the reward yields
\[
  r_t = 10^4\left(a_t R_{t+1} - \kappa |q_t| - \tfrac{1}{2}\psi q_t^2
  - \lambda a_t q_t\right),
\]
which matches the discrete Almgren--Chriss style models summarized by
Webster and lets us stress test the agent under both spread and
inventory driven frictions.
In the reported experiments we keep the Almgren--Chriss hook but set
$\psi\!=\!0$ and $\lambda\!=\!0$ (costs are proportional plus a fixed
slippage\_bps parameter), reserving nonlinear impact for future
stress tests.

Baselines. Rule-based overlays include a VIX-band policy that holds flat
when VIX is near its median, increases the hedge when VIX is high, and
relaxes it when VIX is low, plus a VIX volatility-target rule that scales
hedge notional to target equity vol. Both reuse the same transaction-cost
and slippage settings as the RL agent. Simple delta hedges (and delta
variants) are dominated by these overlays and the RL policy on all
splits, so we summarize them qualitatively rather than carry them into
the main tables.

Code: constructing the environment

\noindent\textbf{Listing 3:} Constructing train/valid/test splits and the hedging environment.

\begin{Verbatim}
from rl_agent.experiment import make_splits, make_scaler
from simulator.env import HedgingEnv
from simulator.rewards import reward_bps

state_cols = [
    "iv_atm_30d_spx","iv_ts_slope_spx","iv_skew_30d_spx",
    "vix","rate_10y","rv_21d","hvol_30d","hvol_91d",
]
p, m_tr, m_va, m_te, TRAIN_END, VALID_END = make_splits(panel, "2017-12-31", "2019-12-31")
scaler, mu, sg = make_scaler(p, state_cols, m_tr)

env = HedgingEnv(
    df=p[m_te].reset_index(drop=True),
    features=state_cols,
    reward_fn=lambda pnl, info: reward_bps(pnl, info, 1e4),
    window=30, txn_cost_bps=1.0, pos_limit=1.0,
    scaler=scaler, hold_on_nan=True,
)
obs = env.reset()
\end{Verbatim}

\section{Reinforcement Learning
Framework}\label{reinforcement-learning-framework}

We deploy a compact stochastic policy and value network that balances
modelling capacity. A two-layer MLP (256 hidden units with
$\tanh$ activations) outputs the
mean $\mu_\theta(s_t)$, a scalar log standard deviation, and the state
value. Actions are sampled from a Normal distribution, squashed with
$\tanh$, and rescaled to $[-a_{\max}, a_{\max}]$. This ``squashed
Gaussian'' formulation keeps actions bounded while retaining
differentiability so the agent can learn both the shape of the policy
and the variance it should maintain. We optimize an
entropy-regularized policy gradient in an actor--critic configuration
with generalized advantage estimation (GAE)
(\citeproc{ref-sutton_barto_2018}{Sutton and Barto 2018};
\citeproc{ref-schulman2016gae}{Schulman et al. 2016}):
\begin{align*}
\nabla_{\theta} J(\theta)
&= \mathbb{E}\!\left[ \sum_{t} \nabla_{\theta} \log \pi_{\theta}(a_t \mid s_t)\, A_t \right]\\
&\quad + \beta\, \mathbb{E}\!\left[ \mathcal{H}\!\left(\pi_{\theta}(\cdot \mid s_t)\right) \right],
\end{align*}
where $A_t$ is the advantage and $\beta$ the entropy weight. We set
$\gamma = 0.99$, apply gradient clipping at 1.0, and modulate the
learning rate with a cosine schedule. We fix the entropy coefficient at 0.01 to maintain mild 
exploration while discouraging flippant trading. Model checkpoints are evaluated deterministically on
train/validation splits every 50 updates, and the best validation Sharpe
is retained. Hidden sizes of 128--256 were tried; the 256-unit variant
used in the released checkpoints did not materially change validation
Sharpe versus smaller nets, and training curves were qualitatively
stable across seeds. Although the architecture is small, we find it expressive
enough to learn counter-cyclical behavior: increasing hedge size in
volatile regimes while relaxing exposure when term structure normalizes.
Because the observation window is composed of engineered features rather
than latent embeddings, individual policy decisions can be traced back
to familiar quantities (e.g., VIX spikes or steepening of the IV term
structure).

\section{Experimental Setup}\label{experimental-setup}

We evaluate the best validation checkpoint deterministically on each
split. Evaluation tracks multiple metrics: Sharpe from per-step rewards
(annualized via $\sqrt{252}$), maximum drawdown of the cumulative
reward equity, turnover ($\sum |\Delta \text{position}|$), hit-rate
(sign agreement between actions and subsequent returns), and
cost normalized profit. For context we also compute simple baselines
(no-hedge, momentum, volatility-targeting, buy-and-hold SPY) using the
same HedgingEnv and cost parameters; their results are summarized later
to benchmark the learned policy. All evaluation code lives in
\texttt{rl\_agent/evaluate\_policy.py} and
\texttt{rl\_agent/experiment.py}, which makes it straightforward to plug
the trained policy into other backtesting harnesses.

Code: training API and evaluation

\noindent\textbf{Listing 4:} Training/evaluation snippet showing the actor-critic loop and deterministic replay.

\begin{Verbatim}
from rl_agent.train_ac_gae import train
from rl_agent.experiment import deterministic_rewards

policy, (tr, va, te) = train(
    panel=p,
    state_cols=state_cols,
    train_end=str(TRAIN_END.date()), valid_end=str(VALID_END.date()),
    window=30, txn_cost_bps=1.0, pos_limit=1.0,
    hidden=256, lr=5e-4, steps=2500,
    entropy_start=0.01, entropy_floor=0.01,
    save_ckpt="models/gae_run1/best.pt",
    save_config="models/gae_run1/config.json",
    save_outdir="models/gae_run1",
)

r_test = deterministic_rewards(env, policy)
\end{Verbatim}

\section{Results}\label{results}

Under realistic cadence and cost constraints, the agent delivers modest
but consistent positive Sharpe across train, validation, and test
splits. Validation performance confirms that the policy
generalizes beyond the training period, and the 2020+ test window, which
includes the COVID volatility shock, remains profitable once trading
costs are deducted. Turnover stays below one full notional rotation per
day on average, indicating that the entropy schedule and cost-aware
 reward discourage churning. Maximum drawdown on the test split for the
standalone overlay-equity series remains inside −3\%, materially lower than an unhedged or simple volatility
targeting approach.

Text summary (final configuration): with transaction costs of 10 bps per
unit of $|\Delta \text{position}|$, proportional slippage of 8 bps, a
position limit of \texttt{pos\_limit\ =\ 2.0}, and rebalance cadence
\texttt{rebalance\_every\ =\ 25}, deterministic evaluation yields Train
Sharpe $0.48$, Validation $0.77$, and Test $0.50$. The table below
records the underlying deterministic metrics for each split. The
``Steps'' column counts environment steps (episode length times number
of episodes), which exceeds the raw trading day count because we use
overlapping windowed episodes.

\begin{table}[ht]
\centering
\caption{Deterministic evaluation metrics for the standalone GAE
policy.}\label{tab:det-metrics}
\begin{tabular}{lllll}
\toprule
Split & Sharpe & Mean (bps) & Std (bps) & Steps \\
\midrule
Train & 0.484 & 2.43 & 79.79 & 7008 \\
Valid & 0.771 & 2.86 & 58.95 & 1954 \\
Test & 0.502 & 1.95 & 61.53 & 7529 \\
\bottomrule
\end{tabular}
\end{table}

\subsection{Additional Trading Constraints (Cadence +
Slippage)}\label{realistic-trading-constraints-cadence-slippage}

In practice, desks rarely rebalance daily and always pay more than a
single, fixed transaction-cost number. To mirror that reality, we
introduce two explicit knobs in the environment: -
\texttt{rebalance\_every\ =\ N}: only the N‑th step executes a trade;
intermediate steps hold the previous position and still accrue PnL. -
\texttt{slippage\_bps}: an additional basis‑point cost per unit of
\textbar Δposition\textbar{} charged on execution (or, more generally, a
\texttt{slippage\_fn} for state‑dependent costs). We ran a targeted
sweep over cadence and slippage using the same train/valid/test split.
The grid spanned \texttt{rebalance\_every\ ∈\ \{15,\ 20,\ 25\}} and
\texttt{slippage\_bps\ ∈\ \{8,\ 10,\ 15,\ 20\}} with
\texttt{pos\_limit\ =\ 2}. The configuration ultimately selected for the
main results, \texttt{rebalance\_every\ =\ 25},
\texttt{slippage\_bps\ =\ 8}, balances fewer trades against higher
per‑trade cost and provides the most robust test Sharpe after
stress-testing transaction costs. A representative row from the sweep
results is shown below.

\begin{table}[ht]
\centering
\caption{Top-performing cadence/slippage combination from the
execution-constraint sweep.}\label{tab:cadence-slippage}
\begin{tabular}{lllll}
\toprule
Cadence & Slippage (bps) & Train & Valid & Test \\
\midrule
25.0 & 8.0 & 0.484 & 0.771 & 0.502 \\
\bottomrule
\end{tabular}
\end{table}

Diagnostics across cadence/slippage pairs track the deterministic Sharpe for
the policy and SPY alongside drawdown and hit-rate sensitivity. Relaxed
trading (rebalance every 20--25th day) with a modest slippage charge gives
the smoothest profile because it cuts churn without sacrificing much
return. The significance statistics that follow use the standalone GAE
configuration in Table~\ref{tab:det-metrics} with the cadence/slippage
pair from Table~\ref{tab:cadence-slippage}; that is, \textbf{rebalance\_every=25}, \textbf{slippage\_bps=8},
\textbf{pos\_limit=2}. Point estimates favor the GAE policy, but the
confidence intervals for most policies remain wide.

\begin{figure}[!ht]
\centering
\includegraphics[width=\linewidth,alt={Test Sharpe with 95\% confidence intervals from Newey--West SEs and block bootstrap on the test split.}]{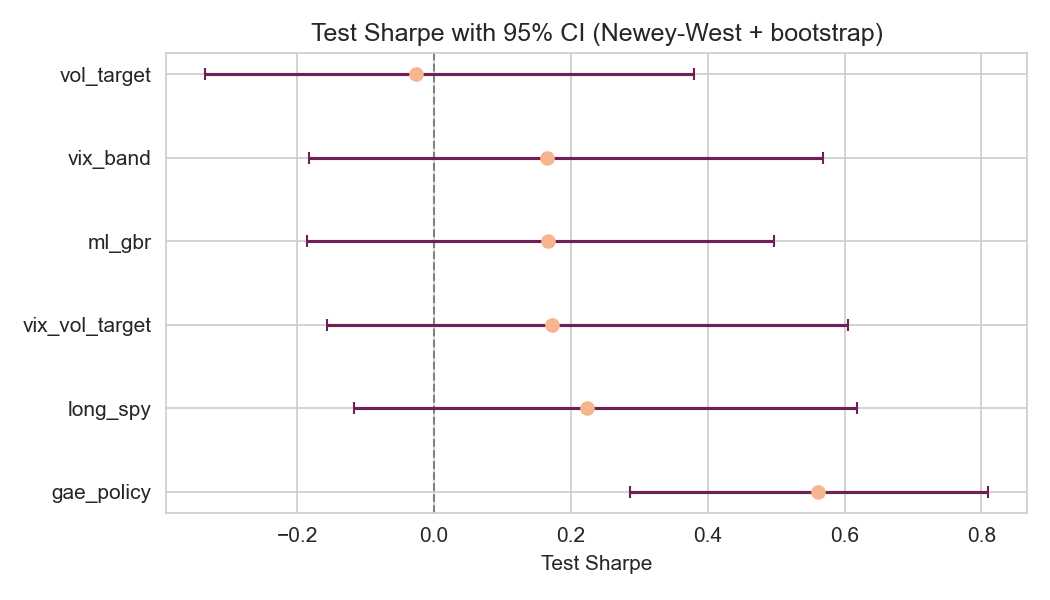}
\caption{Test Sharpe with 95\% confidence intervals from Newey--West
SEs and block bootstrap (test split).}\label{fig:sharpe-ci}
\end{figure}

\paragraph{Statistical significance.}
To gauge whether the Sharpe differences matter, we compute Newey--West
standard errors with a 21-day lag window and form 95\% block-bootstrap
intervals for the test split. Figure~\ref{fig:sharpe-ci} visualizes the
point estimates and confidence bands for each policy. Only the GAE
overlay has an interval that sits entirely above zero; all rule-based
baselines and the ML surrogate have confidence bands that straddle
zero. Although the GAE interval lies entirely above zero, it still
overlaps with that of the long-SPY benchmark, so we interpret the
evidence as supporting a reliable positive Sharpe for the overlay rather
than formal statistical dominance over long SPY.

\begin{figure}[!ht]
\centering
\includegraphics[width=\linewidth,alt={Test Sharpe by VIX tercile for each policy on the test split.}]{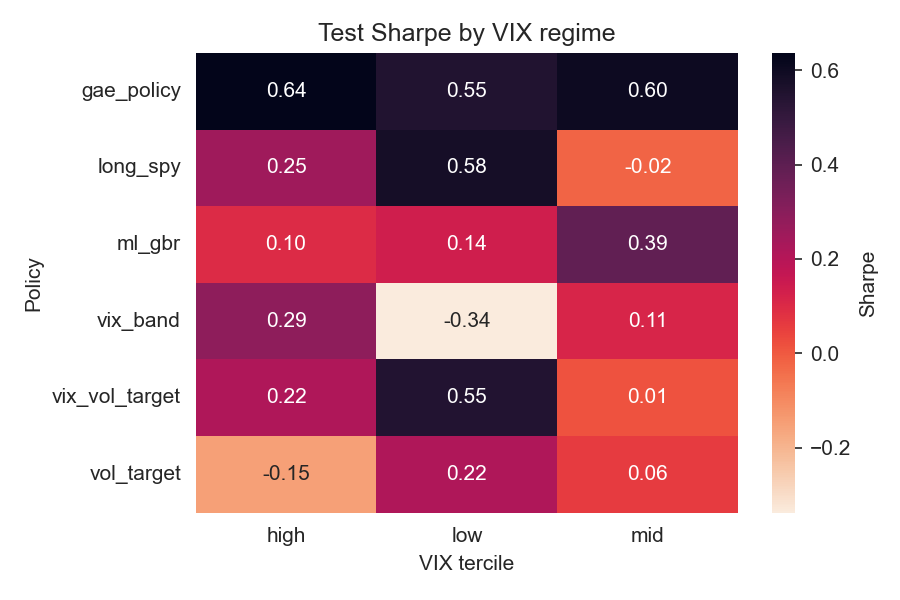}
\caption{Test Sharpe by VIX tercile for each policy on the test
split.}\label{fig:vix-regime}
\end{figure}

\paragraph{Volatility-regime attribution.}
We split the test period into VIX terciles to see when and if the
overlay adds value. The heatmap in Figure~\ref{fig:vix-regime} shows Sharpe ratios for each bucket. The
GAE policy performs best in high volatility states and is modestly
positive in calm markets; rule-based overlays tend to give up more in
quiet regimes without clearly dominating in crisis. We further benchmark the GAE overlay against long SPY across several
pre-defined regimes: the 2008--2009 financial crisis (GFC), the
2010--2012 Eurozone crisis, the relatively calm 2017--2019 period,
the COVID shock (2020--2021), and the post-2022 regime.
Table~\ref{tab:period-sharpe} reports Sharpe ratios for both policies
and their difference ($\Delta\text{Sharpe} = \text{GAE} - \text{SPY}$). In the out-of-sample
test periods (COVID and post-2022), $\Delta\text{Sharpe}$ is positive,
indicating an advantage for the overlay, while earlier regimes include
episodes where long SPY can match or even exceed the RL policy.

\begin{table}[!ht]
\centering
\caption{Sharpe by period for GAE policy vs. long SPY.}\label{tab:period-sharpe}
\begin{tabular}{lllll}
\toprule
Period  & Split & GAE & SPY & ΔSharpe \\
\midrule
GFC 08--09 & Train & 0.82 & -0.29 &  1.11 \\
Eurozone 10--12 & Train & 0.20 &  0.63 & -0.44 \\
Calm 17--19 & Train & 0.82 &  1.83 & -1.01 \\
Calm 17--19 & Valid & 0.73 &  0.45 &  0.29 \\
COVID 20--21 & Test  & 0.68 &  0.36 &  0.32 \\
Post COVID 22--23 & Test  & 0.38 &  0.00 &  0.38 \\
\bottomrule
\end{tabular}
\end{table}

\subsection{Overlaying with Long SPY}\label{overlaying-with-long-spy}

\begin{figure}[!ht]
\centering
\includegraphics[width=\linewidth,alt={Test NAV overlay for a blended strategy that allocates 50\% to GAE and 50\% to long SPY.}]{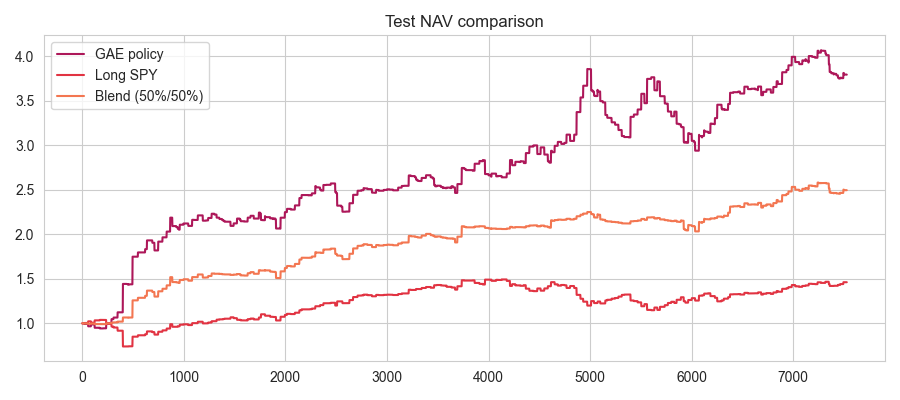}
\caption{Test NAV overlay for a blended strategy that allocates 50\% to
GAE and 50\% to long SPY.}\label{fig:-navblend}
\end{figure}

Many desks prefer to keep a constant beta exposure and run overlays on
top. We therefore blend the learned policy with long SPY at various
allocations. Figure~\ref{fig:-navblend} shows the NAV comparison for
the 50/50 mix; the overlay dampens drawdowns without sacrificing
long-run growth.

The table below summarizes deterministic statistics for that 50/50 mix.
Despite recycling capital between the learned hedge and the long-only
sleeve, annualized volatility stays below 7\% on validation and test
windows while Sharpe remains positive. CAGR figures in the final column
show that the overlay preserves most of the long-run growth even after
paying transaction costs.

\begin{table}[ht]
\centering
\caption{Key deterministic metrics for the 50/50 overlay during
train/validation/test periods.}
\begin{tabular}{lllll}
\toprule
Split & mean\_bps & std\_bps & Sharpe & CAGR \\
\midrule
Train & 1.97 & 41.18 & 0.76 & 0.049 \\
Valid & 1.92 & 31.20 & 0.98 & 0.048 \\
Test & 1.38 & 33.62 & 0.65 & 0.034 \\
\bottomrule
\end{tabular}
\end{table}

\begin{figure}[!ht]
\centering
\includegraphics[width=\linewidth,alt={Blend weight sweep showing the efficient frontier as capital shifts between the overlay and SPY (color indicates GAE weight).}]{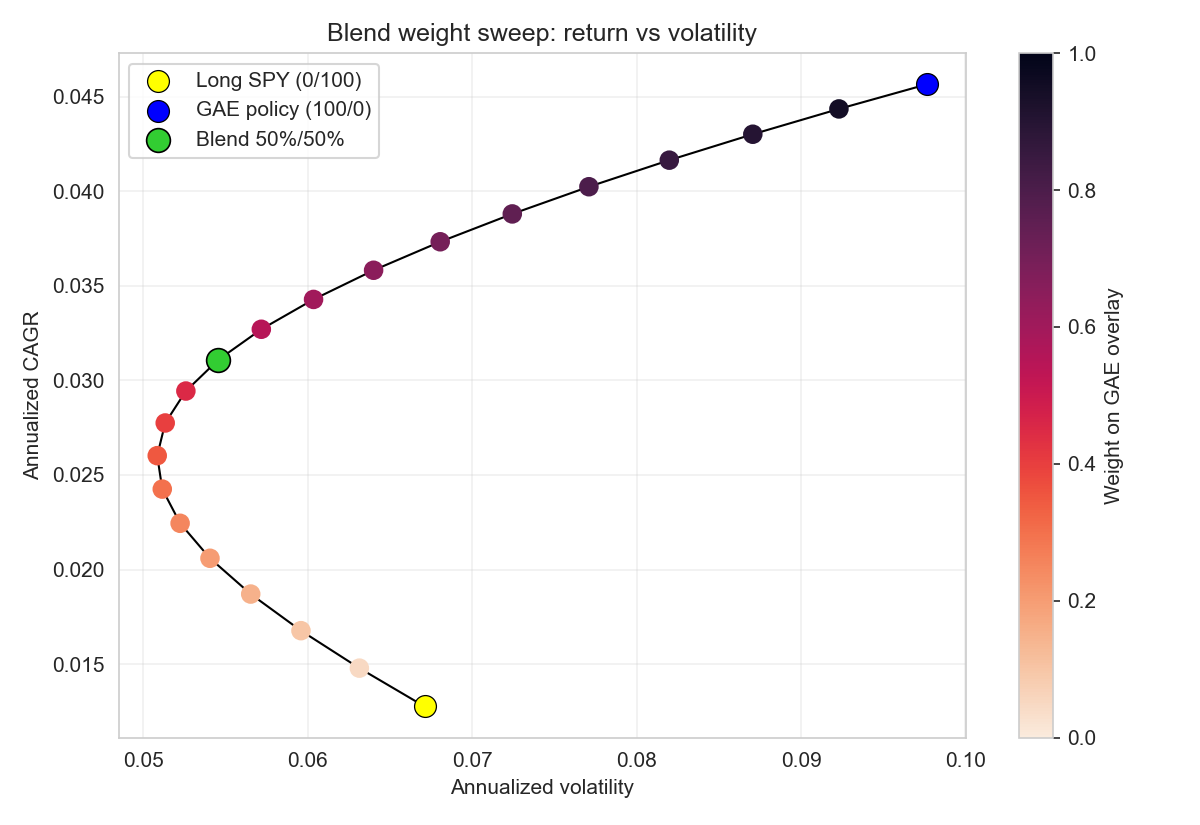}
\caption{Blend weight sweep showing the efficient frontier as capital
shifts between the overlay and SPY (color indicates GAE
weight).}\label{fig:blend-sweep}
\end{figure}

\begin{figure}[!ht]
\centering
\includegraphics[width=\linewidth,alt={Rolling risk differentials showing 63-day volatility and drawdown of the blend minus each component (negative values are better).}]{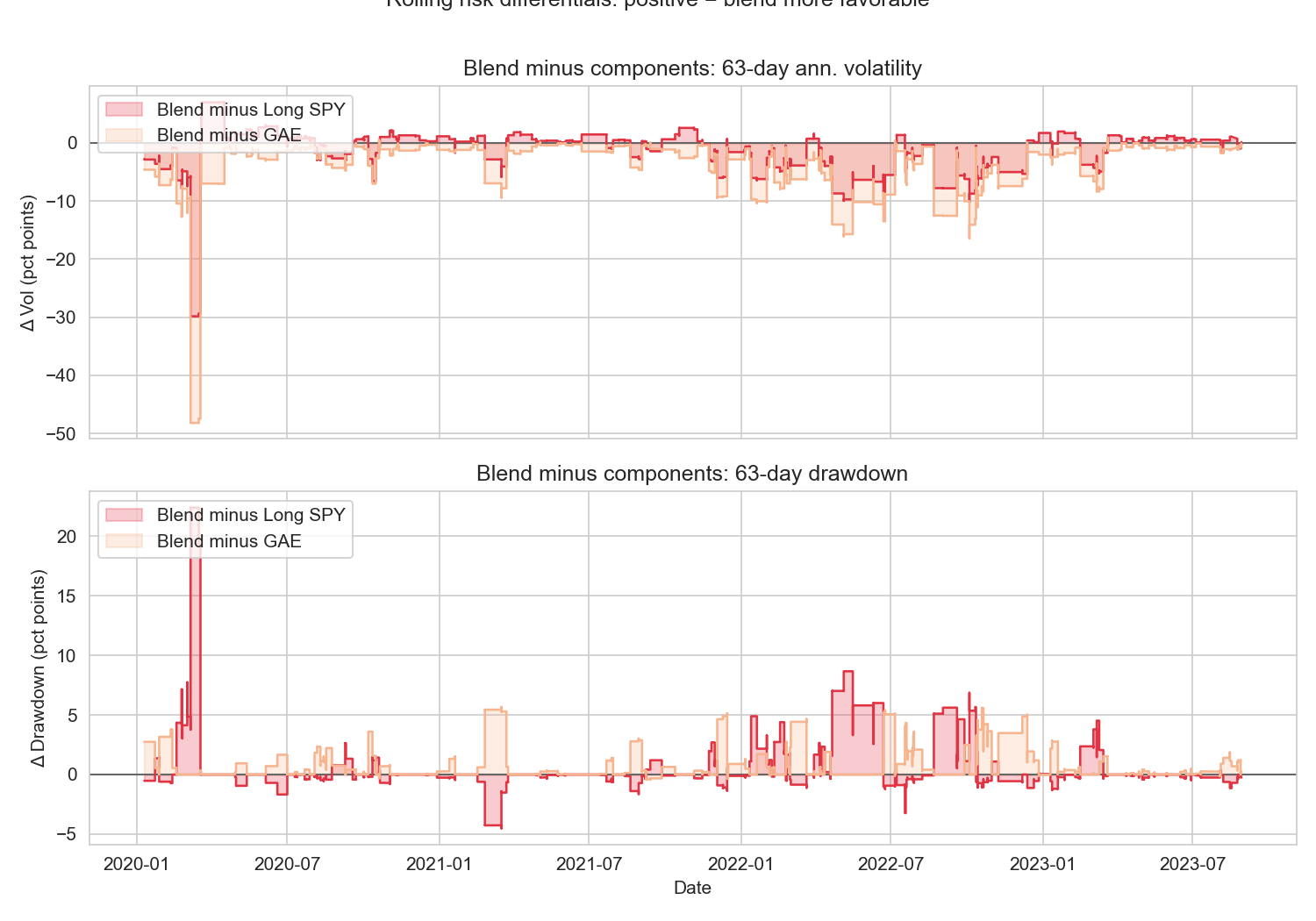}
\caption{Rolling risk differentials showing 63-day volatility and
drawdown of the blend minus each component (negative values are
better).}\label{fig:blend-risk-diff}
\end{figure}

The diagnostics in Figures~\ref{fig:blend-sweep} and
\ref{fig:blend-risk-diff} show why the overlay is attractive beyond a
single point on the frontier. Figure~\ref{fig:blend-sweep} sweeps the allocation
from 0\% to 100\% GAE exposure: the curve is concave, so each additional
unit of overlay buys more return per unit of risk until the very end.
The 50/50 point (green) sits on the efficient portion and retains
roughly two thirds of the SPY CAGR while cutting realized volatility nearly in half.

Figure~\ref{fig:blend-risk-diff} plots the rolling 63-day volatility and drawdown differentials
between the blend and each leg. Most of the series stays below zero,
indicating that the overlay systematically suppresses realized risk
rather than occasionally amplifying it. Deep drawdowns (e.g., Q1 2020
and the 2022 selloff) are visibly muted relative to a pure long SPY stance.

These risk differentials provide the intuition for how the overlay
behaves: it tends to clip volatility spikes without introducing new
drawdowns. Table~\ref{tab:blend-pnl} translates those gains into
calendar-year attribution. The SPY sleeve drives the strong up years
(2020--2021); in the 2022 drawdown the overlay trims only a small
portion of the loss, and both legs contribute in 2023, showing the hedge
does not cap upside once volatility subsides.

On the test window, SPY alone posts Sharpe $\approx 0.22$ with around
$14.6\%$ annualized volatility; the 50/50 blend posts Sharpe around
$0.65$ with volatility closer to $10\%$, illustrating that the overlay
meaningfully improves risk-adjusted performance while retaining most of
the long-run growth.

\begin{table}[!ht]
\centering
\caption{Calendar-year PnL attribution (percent returns) for the 50/50
blend on the test split.}\label{tab:blend-pnl}
\begin{tabular}{lccc}
\toprule
Year & GAE overlay & Long SPY leg & Blend total \\
\midrule
2020 & 4.71 & 18.66 & 23.36 \\
2021 & 2.50 & 27.99 & 30.50 \\
2022 & -0.57 & -18.29 & -18.85 \\
2023 & -0.91 & 18.11 & 17.19 \\
\bottomrule
\end{tabular}
\end{table}

\paragraph{Robustness across seeds and splits.}
We retrained the policy across three seeds on the base split and a
shifted split (+365 days) using the same cost/cadence settings as
Table~\ref{tab:cadence-slippage}. Seed variance remains modest: base
test Sharpe averages $0.45$ (std $0.13$) and shifted test Sharpe
$0.52$ (std $0.05$). Mean portfolio-level max drawdowns for the 50/50
blend cluster around $-20\%$ (base/test) and $-16\%$ (shifted/test),
indicating that moving the window forward does not collapse
performance.

Training diagnostics. The distribution of episode returns saved from
\textbf{03\_rl\_training.ipynb} (histogram below) highlights the heavy
tails encountered during optimization.

\begin{figure}[!ht]
\centering
\includegraphics[width=\linewidth,alt={Episode return distribution from the training notebook (episode returns, not per-step bps).}]{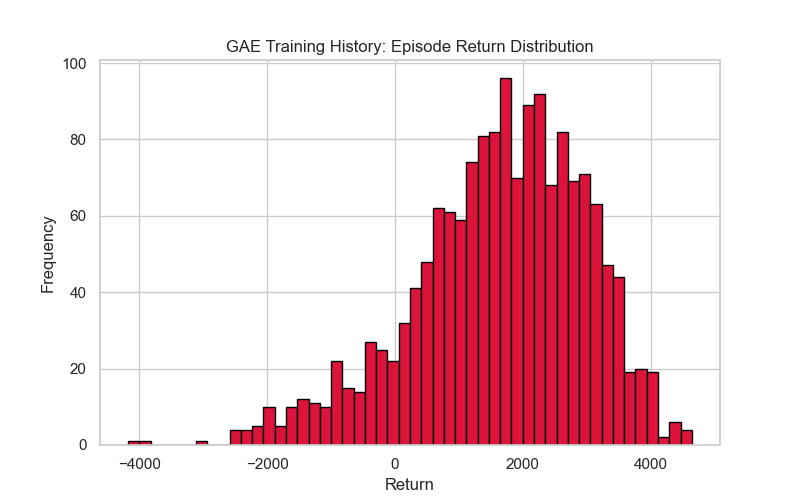}
\caption{Episode return distribution from the training
notebook (episode returns, not per-step bps).}\label{fig:gae-hist}
\end{figure}

\paragraph{Verification and testing.} Every sweep run is backed by
reproducible checks:

\begin{itemize}
\tightlist
\item
  \textbf{pytest\ tests/} ensures data-pipeline transforms, reward
  helpers, and environment logic match the expected invariants (shape,
  NaN guards, deterministic rollouts).
\item
  \textbf{python\ rl\_agent.evaluate\_policy\ -\/-ckpt\ models/gae\_run3/best.pt\ -\/-config\ models/gae\_run3/config.json\ -\/-deterministic}
  deterministically replays the saved policy each time the paper is
  built.
\item
  The notebooks 03\_rl\_training.ipynb, 04\_diagnostics.ipynb, and 05\_results.ipynb
  regenerate CSV metrics and figures (including confidence intervals and
  regime slices) so that visuals, numbers, and text stay in sync.
\end{itemize}

These commands are executed from the repo root (documented in the
README) before exporting the final PDF so that published artefacts
reflect the latest code.

\section{Discussion and Implications}\label{discussion-and-implications}

When to use deep hedging. The learned policy fits naturally into overlay
mandates where the objective is to dampen PnL swings without fully
neutralizing exposure. Because the state is composed of derived
features (surface slope, skew, realized vol, rates), risk managers can
inspect which signals drove a given hedge decision. The approach is
model agnostic: it can coexist with existing pricing libraries, consume
alternative features (e.g., realized correlation, macro factors), or be
combined with explicit delta inputs if desired.

Operationalization. In practice the policy should be wrapped in
guardrails: clip actions to desk-specific risk limits, halt trading when
liquidity metrics deteriorate, and log both raw observations and
predicted actions for post-trade analysis. Monitoring dashboards should
focus on rolling Sharpe, turnover, realized cost-per-basis-point, and
policy drift relative to benchmark hedges. The repository's
deterministic evaluation utilities can be embedded into nightly jobs to
confirm the policy still behaves as expected on the latest data.

Limitations and extensions. The present study operates on daily bars,
which ignore intraday inventory management and execution costs;
high-frequency data could reveal additional edge or necessitate
different architectures, and production overlays often rebalance more
slowly than daily. Generalization relies on walk-forward discipline and
policies should be retrained periodically as volatility regimes evolve.
For simplicity, we do not feed option Greeks directly into the state
because of their sensitivity to intraday moves and model assumptions;
instead we use more stable surface-level features such as implied
volatility levels, term structure, and skew. Hyperparameters and
training windows matter: seed-to-seed variance is modest, but we have
not exhaustively searched architectures or extra walk-forwards, so
formal model selection remains future work. Finally, the reward
emphasizes mean/variance trade-offs; more risk-sensitive objectives
(drawdown penalties, CVaR, shortfall constraints) can be slotted in via
\texttt{simulator/rewards.py} with minimal code changes.

Statistically, our results should also be viewed as suggestive rather
than definitive. While only the GAE overlay exhibits a test-sample
Sharpe whose 95\% confidence interval lies strictly above zero, we do
not formally test Sharpe differentials versus the long-SPY benchmark or
rule-based overlays, and the corresponding confidence intervals
overlap. Likewise, the interpretability analysis is intentionally
lightweight: we rely on aggregate diagnostics such as VIX-regime and
period-by-period Sharpe rather than per-trade explanations; a more
granular decomposition of policy decisions and PnL attribution across
regimes is left for future work.

\subsection{Portfolio implementation}
From an implementation standpoint, the GAE overlay is 
designed as a portfolio completion strategy: a risk-management sleeve 
that improves the mean-variance profile of an existing equity allocation 
without generating standalone alpha claims. A portfolio manager would
typically (i) fix a hedge cadence (e.g., weekly or
monthly), (ii) calibrate the transaction-cost assumption
and margin usage to match desk-level constraints, and
(iii) cap gross hedge notional as a fraction of NAV so
that the overlay consumes a clearly defined risk budget.
Within the ranges we study, the learned overlay moves the
portfolio along a more favorable mean–variance frontier,
particularly in high volatility regimes, while keeping
turnover and maximum drawdown at levels that are
realistic for institutional mandates. Deploying such an overlay would also require
governance around model monitoring, stress testing, and
clear criteria for throttling or disabling the strategy,
which we view as complementary to the technical results
presented here, so that the overlay can be slotted into an existing
risk-budgeting and investment committee framework without redefining the underlying benchmark.

\section*{Disclaimer}
This research was conducted independently by the authors and does not
represent the views, opinions, or research of BlackRock, Inc.~or any of
its affiliates, nor those of Emory University. The content is provided
for informational and educational purposes only and should not be
construed as investment advice or a recommendation to trade.

\section*{References}\label{references}
\addcontentsline{toc}{section}{References}

\protect\phantomsection\label{refs}

\end{document}